\documentclass[reprint,showpacs,preprintnumbers,superscriptaddress,nofootinbib,amsmath,amssymb,aps,prl,floatfix]{article}
\pdfoutput=1
\usepackage{jcappub}
\usepackage{graphicx}
\usepackage{url}
\usepackage[center]{subfigure}
\usepackage{xspace}
\usepackage{footnote}

\allowdisplaybreaks	

\newcommand{\camb}{\texttt{CAMB}\xspace}
\newcommand{\cosmomc}{\texttt{CosmoMC}\xspace}
\newcommand{\hub}{\ensuremath{\mathcal{H}}}
\newcommand{\lcdm}{$\Lambda$CDM}
\newcommand{\wla}{\ensuremath{w_\Lambda}}
\newcommand{\wlae}{\ensuremath{w_\Lambda^{\rm{eff}}}}
\newcommand{\rhodm}{\ensuremath{\rho_{\rm{DM}}}}
\newcommand{\rhode}{\ensuremath{\rho_\Lambda}}

\newcommand{\neff}{\ensuremath{N_{\rm{eff}}}}
\newcommand{\meff}[1]{\ensuremath{m_{#1}^{\rm{eff}}}}
\newcommand{\nus}{\ensuremath{+\nu_s}}
\newcommand{\logA}{\ensuremath{\log(10^{10}A_s)}}

\newcommand{\Hou}{\ensuremath{\, \text{Km s}^{-1}\text{ Mpc}^{-1}}}

\newlength{\lenerrbars}
\newlength{\lencontours}
\newlength{\lencls}
\begin{document}


\title{Constraints on the Coupling between Dark Energy and Dark Matter from CMB data}

\author[a,b,c]{R. Murgia}
\emailAdd{riccardo.murgia@sissa.it}

\author[a,b]{S. Gariazzo}
\emailAdd{gariazzo@to.infn.it}

\author[a,b]{N. Fornengo}
\emailAdd{fornengo@to.infn.it}

\affiliation[a]{Department of Physics, University of Torino, Via P. Giuria 1, I--10125 Torino, Italy}
\affiliation[b]{INFN, Sezione di Torino, Via P. Giuria 1, I--10125 Torino, Italy}
\affiliation[c]{SISSA, via Bonomea 265, 34136 Trieste, Italy}


\abstract{
We investigate a phenomenological non-gravitational coupling between dark energy and dark matter, where the
interaction in the dark sector is parameterized as an energy transfer either from dark matter to dark energy or the opposite.
The models are constrained by a whole host of updated cosmological data: cosmic microwave background temperature anisotropies and polarization, high-redshift supernovae, baryon acoustic oscillations, redshift space distortions and gravitational lensing. 
Both models are found to be compatible with all cosmological observables, but in the case where dark matter decays into dark energy, the tension with the independent determinations of $H_0$ and  $\sigma_8$, already present for standard cosmology, increases: this model in fact predicts lower $H_0$ and higher $\sigma_8$,  mostly as a consequence of the higher amount of dark matter at early times, leading to a stronger clustering during the evolution. Instead, when dark matter is fed by dark energy, the reconstructed values of $H_0$ and  $\sigma_8$ nicely agree with their local determinations, with a full reconciliation between high- and low-redshift observations. A non-zero coupling between dark energy and dark matter, with an energy flow from the former to the latter, appears therefore to be in better agreement with cosmological data.
}


\maketitle

\section{Introduction}
\label{sec:intro}
Cosmological observables, among which the cosmic microwave background (CMB) radiation temperature ani\-so\-tro\-pies represents one of the most important pillars of modern cosmology, providing a consistent and convincing picture that  the Universe is largely dominated by ``dark'' components, whose presence is felt only through their gravitational effect. Specifically, the recent results of the Planck Collaboration \cite{Adam:2015rua}  show that baryonic matter forms only about 5\% of the total energy density today, while 26\% is accounted for by dark matter (DM), which clusters forming cosmic structures, and the remaining 69\% is in the form of a diffuse component, the dark energy (DE), responsible of the accelerated expansion of the Universe, being the CMB radiation (and relic neutrinos) negligible in the energy budget today. While Planck is currently providing the most precise determination of cosmological parameters, a whole set of independent cosmological observations (among which high-redshift supernovae, baryon acoustic oscillations, redshift space distortions and gravitational lensing) consistently contribute to what has emerged as a successful interpretation of the behavior of our Universe, namely the $\Lambda$CDM model. Some tension exists, though, on the cosmological and local determination of the Hubble parameter $H_0$ and on $\sigma_8$, the amplitude of the linear power spectrum on the scale of 8 $h^{-1}$ Mpc
(being $h$ the usual reduced Hubble constant, $h= H_0/(100$ km s$^{-1}$ Mpc$^{-1}$)), as discussed for example in Ref.~\cite{Ade:2015xua}. 

The nature of DM and DE is still unknown and while DM can be naturally interpreted as being formed by a new type of elementary particle (typically heavy, in the standard WIMP paradigm, but not necessarily, like in the case of axions or axion-like particles), a clear view on the solution of DE is still lacking. The simplest interpretation of DE is in the form of a cosmological constant $\Lambda$: in standard cosmology, based on general relativity, the cosmological constant can be seen as a fluid endowed with equation of state (EoS) $p_\Lambda=\wla\rhode$, 
with $\wla=-1$ ($p_\Lambda$ and $\rhode$ being the pressure and the energy density of DE, respectively): this allows to explain the current accelerated phase of the Universe and what we call DE.

However, at the fundamental level DE could be more complex than just a new fundamental constant. A more general
view on the DE problem is obtained by seeing the issue from a dynamical side: a (perfectly or almost perfectly) homogeneous scalar field can act as a cosmological constant if its potential energy dominates over its kinetic energy (and this energy density largely dominates the Universe) \cite{Kodama:1985bj,Dunsby:1991xk}. This mechanism has been studied in details, also in connection with the so-called ``coincidence-problem'' (why DM and DE are so close today in value, having such a different cosmological evolutionary behavior) and many models have been devised to investigate the ability of a dynamical scalar field to explain the cosmological observations. For the purposes of this paper, the main point is that DE can therefore be seen as an effect due to the presence of a dynamical scalar field in a specific phase of its cosmological evolution.

Under the assumption that DE is due to a scalar field, the most economic approach is that the two sectors, i.e.\ DM and DE, do not interact other than gravitationally. However, the two sectors might well have some form of non-gravitational interactions, with relevant implications both at the fundamental level (their true nature) and at the cosmological level, since cosmological observables can be affected by the presence of an interaction between DM and DE (see e.g.\
Refs.~\cite{Friedman:1991dj,Gradwohl:1992ue,Wetterich:1994bg,Amendola:1999er,Amendola:2003wa,Pettorino:2008ez,Valiviita:2008iv,
He:2008tn,Gavela:2009cy,He:2009mz,Abdalla:2014cla,Salvatelli:2014zta,Lee:2006za,Xue:2014kna,Faraoni:2014vra,Koivisto:2015qua} 
and the review \cite{Bolotin:2013jpa}). At the same time, cosmological and astrophysical observations can be proficiently used to constrain any form of coupling between the two dark sectors.

This new interaction can be phenomenologically introduced in various ways (see e.g.\ Ref.~\cite{Koyama:2009gd} for a classification): in our analysis, which follows similar approaches
\cite{He:2008tn,Gavela:2009cy,He:2009mz,Abdalla:2014cla,He:2008si,Jackson:2009mz,Salvatelli:2013wra, Costa:2013sva,Gavela:2010tm,He:2010im}, 
we phenomenologically parameterize the coupling between DM and DE through an energy transfer from one sector to the other. This can be expressed through the non-conservation of their stress-energy tensors $T^{\mu\nu}_i$ ($i=$DM, DE). In this approach, the stress-energy tensors of DM and DE are not  separately conserved (while the total stress-energy tensors of DM and DE is), and we parameterize this as:
\begin{subequations}
\begin{eqnarray}
 \nabla_\mu T^{\mu\nu}_{\rm{DM}} & = & +Q u^\nu_{\rm{DM}}/a\label{eq:stressenergyTDM}\,,\\
 \nabla_\mu T^{\mu\nu}_{\rm{DE}} & = & -Q u^\nu_{\rm{DM}}/a\label{eq:stressenergyTDE}\,,
\end{eqnarray}
\end{subequations}
where the coefficient $Q$ encodes the interaction between the two sectors, 
$u^\nu_{\rm{DM}}$ is the dark matter four-velocity and 
$a$ is the time-dependent scale factor of the Universe
\cite{He:2008tn,Gavela:2009cy,He:2009mz,Abdalla:2014cla,He:2008si,Jackson:2009mz,Salvatelli:2013wra, Costa:2013sva,Gavela:2010tm,He:2010im}. 
The ensuing evolution equations for the DM and DE energy densities are therefore\footnote{We use the subscript $\Lambda$ also for DE, since the equations are the same. 
The difference is that the cosmological constant $\Lambda$ has a constant EoS parameter $\wla=-1$, 
while in general DE can have $\wla\neq-1$. More generally, DE can also have a dynamical $\wla$. In our analysis, will consider only the case of a (effectively) constant EoS parameter.}:
\begin{subequations}\label{eq:dmdecou}
  \begin{eqnarray}
 \dot{\rho}_{\rm{DM}}+3\hub\rhodm &=& +Q \;, \label{eq:dmdecou_dm}\\
 \dot{\rhode}+3\hub(1+\wla)\rhode &=& -Q \,.  \label{eq:dmdecou_de}
  \end{eqnarray}
\end{subequations}
If $Q>0$ the energy transfer is from DE to DM and DE decays into DM, while 
if $Q<0$ the energy flux has the opposite direction and DM decays into DE.

Several interaction models have been proposed in the literature,
where the role of Coupled Dark Energy (CDE in the following)
is played by a scalar field (see e.g.\ Refs.~\cite{Amendola:1999er,Amendola:2006dg,Valiviita:2008iv,He:2008si,CalderaCabral:2009ja,Jackson:2009mz,Pavan:2011xn}).
In this work we will not focus on specific theoretical frameworks
that give origin to a CDE scenario. 
We instead use a phenomenological approach and we study one specific model, investigated in the past in Refs.  \cite{He:2008tn,Gavela:2009cy,He:2009mz,Abdalla:2014cla,He:2008si,Jackson:2009mz,Salvatelli:2013wra, Costa:2013sva,Gavela:2010tm,He:2010im}, where the coupling term is proportional to the DE density:
\begin{equation}
\label{eq:coupl}
Q=\xi \hub \rhode\,,
\end{equation}
where $\xi$ is a dimensionless coupling parameter.
In this approach the interaction is spatially-independent (being so the DE density) and
the time dependence of the interaction rate is governed by the Hubble rate 
$\hub = \dot{a}/a$ \cite{Zimdahl:2001ar,Salvatelli:2013wra, Costa:2013sva}. For $\xi=0$ we recover the uncoupled case
of standard cosmology.
As we will show below, this model offers the possibility to loosen the tension between the local and cosmological determinations of the Hubble constant $H_0$ and the matter fluctuations parameter $\sigma_8$.
Further models will be studied elsewhere.

In the following we will test this CDE model against cosmological observables and derive bounds on the relevant model parameters, that in our approach are $\wla$ and $\xi$. We will also discuss whether the ensuing results help in alleviating the tension on the determination of $H_0$ and $\sigma_8$ which arises from high and low redshift cosmological observables.

The outline of this work is the following:
in Section~\ref{sec:param} we describe our parameterization for the \lcdm~model 
and its extension when there is a coupling between DE and DM.
In Section~\ref{sec:data} we present the cosmological data used to test our model.
In Section~\ref{sec:results} we show and discuss our results.
In Section~\ref{sec:sterilenuDM} we extend the analysis to include sterile neutrinos: in this case DM is partly composed of a fraction interacting with DE and a stable fraction, represented by a sterile neutrino.
Finally, we summarize our conclusions in Section~\ref{sec:disc}.

\section{Parameterization}
\label{sec:param}
In this Section we introduce the parameterization we use to perform our cosmological analyses.
Subsection~\ref{ssec:lcdm} briefly recall the standard \lcdm~model, which is our baseline model to which we will compare the results of the CDE modeling. In Subsection~\ref{ssec:cde} we then discuss in more detail the modifications that arise in the cosmological model by the presence of a coupling between DM and DE: this occurs both at the background and perturbations level.

\subsection{Baseline model}
\label{ssec:lcdm}
Our baseline model is the well studied standard \lcdm~model, 
where $\Lambda$ indicates the cosmological constant and CDM stands for Cold Dark Matter.
The model can be described using six cosmological parameters:
the present baryon density $\Omega_bh^2$;
the present CDM density $\Omega_ch^2$;
the ratio of the sound horizon to the angular diameter distance at decoupling $\theta$;
the optical depth at reionization $\tau$;
the amplitude $A_s$ and
the spectral index $n_s$ 
of the primordial power spectrum of scalar perturbations, taken at the pivot scale $k=0.05$ Mpc$^{-1}$.
The DE density is a derived parameter, and we consider the case of a flat Universe.
The current value of the Hubble parameter $H_0$ and
the root-mean-square fluctuations in total matter in a sphere of $8h^{-1}$~Mpc radius,
$\sigma_8$, 
are derived
parameters as well.

Concerning neutrinos, we will perform two type of analysis: in the first one we fix the sum of the neutrino masses $\sum m_\nu$ to the minimal value allowed by the neutrino oscillations 
($\sum m_\nu=0.06$~eV) 
and the effective number of relativistic species $\neff$ 
to the standard value $\neff^{\rm{sm}}=3.046$ \cite{Mangano:2005cc}.
In Section~\ref{sec:sterilenuDM} we will extend the scenario to the presence of one additional sterile neutrino,
that plays the role of a stable fraction of DM.

For the \lcdm~parameters we adopt flat priors in the ranges listed in Tab.~\ref{tab:priorslcdm}.
  
\begin{savenotes}
\begin{table}[t]
\begin{center}
\begin{tabular}{|c|c|c|}
  \hline
  Parameter		& Prior		\\	\hline	
  $\Omega_bh^2$ 	& [0.005, 0.1]	\\
  $\Omega_ch^2$ 	& [0.001, 0.5]	\\
  $100\theta$ 		& [0.5, 10]	\\ 	
  $\tau$  		& [0.01, 0.8]	\\ 
  $\logA$ 		& [2.7, 4]	\\ 	
  $n_s$ 		& [0.9, 1.1]	\\ 	\hline
  $\sum m_{\nu}$	& 0.06 eV	\\
  $N_{\nu}$		& 3.046 \cite{Mangano:2005cc}	\\ 	\hline
  $H_0$~[\Hou]		& [20,100]	\\	\hline
\end{tabular}
\caption{Priors and constraints on the parameters adopted in the analysis.
The first six lines refer to priors for the cosmological \lcdm~free parameters,  flat in the listed intervals.
The total neutrino mass and effective number of neutrinos are kept fixed and set at the minimal value allowed by neutrino oscillations in the case of Normal Hierarchy.
$H_0$ is a derived parameter in our analysis, 
but models that
predict extreme and very unlikely values for the Hubble rate today are rejected.
We assume a flat Universe.
}
\label{tab:priorslcdm}
\end{center}
\end{table}
\end{savenotes}

\subsection{Coupling between DE and DM}
\label{ssec:cde}

As introduced in Section \ref{sec:intro}, the coupling between the two dark components of the Universe is
introduced through a phenomenological coupling parameterized through 
the term $Q$ written in Eq.~\eqref{eq:coupl}. 
%
%
The evolution of the DM and DE densities can be easily obtained by solving
Eqs.~\eqref{eq:dmdecou_dm} and \eqref{eq:dmdecou_de} \cite{Izquierdo:2010qy,Gavela:2009cy,Costa:2013sva}:
\begin{subequations}
\label{eq:coupl_bg}
  \begin{eqnarray}
  \rhodm &=& \rho_{\rm{DM}}^0\,a^{-3} + {\rhode^0 a^{-3} \Bigg[\frac{\xi}{3\wla+\xi}
  \big(1-a^{-3\wla-\xi}\big)\Bigg]}\label{eq:coupl_bgDM}\,,\\
  \rhode &=& \rhode^0\, a^{-3(\wla+1)-\xi}\,,\label{eq:coupl_bgDE}
  \end{eqnarray}
\end{subequations}
where $\rho_{i}^0$ ($i=$DM, DE) is the energy density of the species $i$ today.
We recall that $\xi<0$ corresponds to an energy flux from DM to DE, with DM decaying into DE, 
whereas $\xi>0$ corresponds to an energy flux from DE to DM, with DE decaying into DM.
In the following we will refer to the former case as Model 1 (MOD1) and to the latter case as Model 2 (MOD2) 
for sake of brevity.
From Eq.~\eqref{eq:coupl_bgDE} we see that DE obeys an effective equation of state parameter given by 
$\wlae=\wla+\xi/3$: this allows to express Eq.~\eqref{eq:coupl_bgDE} also as $\rhode = \rhode^0\, a^{-3(\wlae+1)}$.

In the presence of the coupling term of Eq.~\eqref{eq:coupl}, 
the interaction model does not suffer gravitational instabilities 
if $\wla \neq -1$~\cite{Valiviita:2008iv,He:2008si}:
for this reason we will consider a constant $\wla \neq -1$ when $\xi\neq0$.
Early time instabilities can however arise also when $\wla\neq-1$ if the coupling is strong \cite{Gavela:2009cy}:
in particular the instability is not present if $\xi$ and $\wla+1$ have opposite sign,
but it can be generated when the two quantities have the same sign.
In order to avoid these instabilities, we will consider only values $\wla>-1$ for MOD1 when $\xi<0$, and
values $\wla<-1$ for MOD2, when $\xi>0$.
It is worthwhile to note that in the latter case 
the DM energy density can assume unphysical negative values in the past 
for particular combinations of $\wla$ and $\xi$ (Eq.~\eqref{eq:coupl_bgDM}),
while the DE energy density is always positive (Eq.~\eqref{eq:coupl_bgDE}).
To avoid unphysical values of $\rhodm$, we must therefore impose $\xi \lesssim -\wla$:
this is automatic for $\xi<0$ (MOD1) unless $\wla$ assumes positive values, 
but this does not occur since accelerated expansion of the Universe at late times requires $\wla<-1/3$.
For MOD2, instead, we impose the prior $0\leq\xi\leq0.5$: we will nevertheless find that
the larger values of $\xi$ in this interval are disfavored by our analyses.

From Eq.~\eqref{eq:coupl_bgDE} we also note that $\rhode$ increases with the scale factor if~$\wla < (-1 -{\xi}/{3})$:
in this region of the parameter space DE has an effective phantom behavior, that is an unbounded increase of $\rhode$ at future times. 
The effective phantom behavior can occur in both MOD1 and MOD2.
Even when $\wla > -1$ and $\xi<0$ (MOD1) the phantom regime can be present since, when the scale factor $a$ increases, 
instead of following a decreasing behavior driven by $\wla > -1$, 
$\rhode$ can be fed by the energy transfer from DM to DE.
This effective behavior, however, has the advantage of being free from the instabilities 
that can occur for a true phantom dark energy \cite{Huey:2004qv,Das:2005yj,Boriero:2015loa}. 

The instabilities that we listed are the consequence of the $\wla=-1$ crossing.
A way to circumvent this problem is to adopt a parameterized post-Friedmann approach (PPF) \cite{Hu:2007pj,Hu:2008zd,Fang:2008sn}. This method allows to seamlessly move through the crossing.
The PPF approach has been extended to deal with an additional coupling
between DM and DE in Ref.~\cite{Li:2014eha}. In the following we will adopt the canonical approach, which is more direct in formulation even though it requires us to separate the two regimes around the crossing point.

Looking at Eqs.~\eqref{eq:coupl_bgDE} and \eqref{eq:coupl_bgDM}, we notice that 
it is difficult to disentangle the effects of the DE EoS parameter $\wla$
from the coupling $\xi$ by only studying the background evolution: 
we must include the perturbation evolution equations, which are also affected by the DM/DE coupling.
The perturbation equations in the linear regime can be expressed in the synchronous gauge \cite{Costa:2013sva} as:
\begin{subequations}
\label{eq:pertcou}
\begin{eqnarray}
\dot{\delta}_{\rm{DM}}
& = 
&-\left(kv_{\rm{DM}}+\frac{\dot{h}}{2}\right)+ {\xi \hub \frac{\rhode}{\rhodm}(\delta_\Lambda-\delta_{\rm{DM}})}
\,;\label{eq:pertcouDMd}\\ 
\dot{v}_{\rm{DM}}& = &-\hub v_{\rm{DM}}\left(1+{\xi\frac{\rhode}{\rhodm}}\right)
\,;\label{eq:pertcouDMv}\\ 
\dot{\delta}_{\Lambda}
& = 
&-(1+\wla)\left(kv_\Lambda+\frac{\dot{h}}{2}\right)
-3\hub(1-\wla)\cdot \left(\delta_\Lambda \hub(3(1+\wla)+{\xi})\frac{v_\Lambda}{k}\right)
\,;\label{eq:pertcouDEd}\\ 
\dot{v}_{\Lambda}& = &-2\hub\left(1+{\frac{\xi}{1+\wla}}\right)v_\Lambda+k\frac{\delta_\Lambda}{1+\wla}
\,;\label{eq:pertcouDEv}
\end{eqnarray}
\end{subequations}
\looseness=-1
where $h=6\phi$ is the synchronous gauge metric perturbation 
and the DM peculiar velocity ${v}_{\rm{DM}}$ is fixed to zero using the gauge freedom.
Moreover, the DE sound speed is fixed: $c_{s,\Lambda}=1$.
We adopt adiabatic initial conditions for the CDE component \cite{Valiviita:2008iv,He:2008si,Gavela:2010tm}
as for all the other cosmological constituents \cite{Ma:1995ey}.

For our cosmological analyses, we implemented the modified relevant equations 
into the numerical Boltzmann solver \camb~\cite{Lewis:1999bs}
and we modified the Markov Chain Monte Carlo (MCMC) code \cosmomc~\cite{Lewis:2002ah} 
in order to include $\xi$ as an additional parameter.
The parameters specific to the DM/DE coupling sector are $\xi$ and $\wla$: they have been varied inside the intervals listed in Tab.~\ref{tab:priorscde}, and we assumed flat  priors. The range of these intervals are taken according to the discussion listed above.

\begin{figure}[t]
  \centering
  \includegraphics[page=1,width=\lencls]{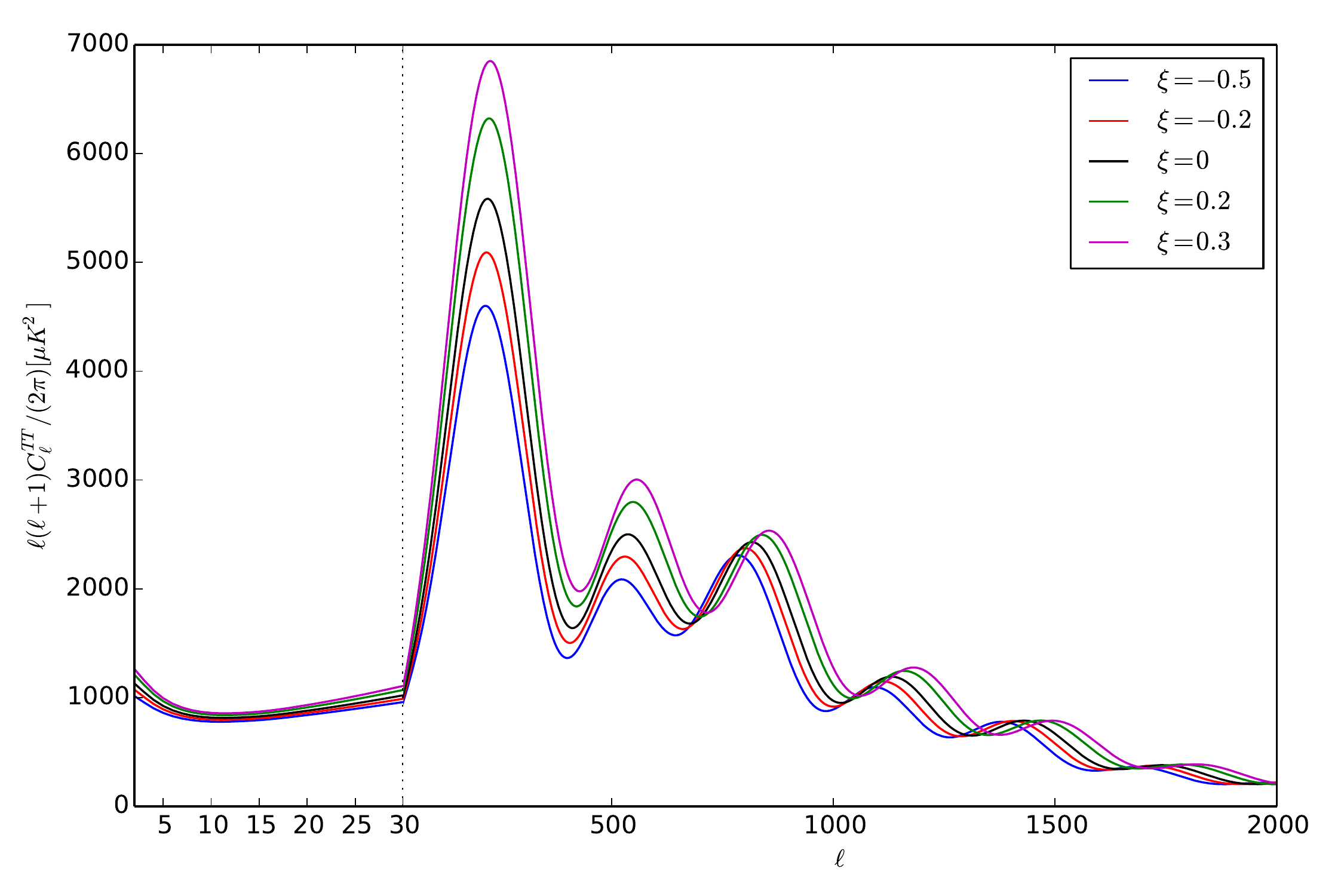}
  \includegraphics[page=2,width=\lencls]{images/cls.pdf}
  \caption{
  Dependence of the CMB angular power spectrum $C_l$ on two
  cosmological parameters:
  the DM-DE coupling strength $\xi$ (upper panel);
  the DM energy density today $\Omega_c h^2$
  (lower panel).
  All the other parameters are kept fixed.
  The black curve is the same in the different panels. The plot shows the degeneracy between $\xi$ and $\Omega_c h^2$.
  }
  \label{fig:cls}
\end{figure}

Before discussing the results of the analysis in details, we comment on some expected effects induced by the DM/DE coupling. First of all a degeneracy between the coupling parameter $\xi$ and the DM density today $\Omega_c h^2$ is expected, due to the conversion of DM into DE (or vice-versa) that reduces (increases) the DM abundance at different times: this impacts the CMB power spectrum since it alters the matter-radiation equality epoch and it changes the matter potentials at CMB decoupling. Fig.~\ref{fig:cls} shows this approximate degeneracy. In the upper panel, where 
$\Omega_c h^2$ is kept fixed, the plot outlines the effect of the DM/DE coupling in the angular power spectrum; in the 
lower panel, instead, the DM/DE energy parameter is fixed to $\xi=0$ (i.e.\ standard $\Lambda$CDM cosmology) and the DM density is varied. A positive coupling (which implies a DE $\rightarrow$ DM transfer) amplifies and shifts the acoustic peaks in a way similar to a Universe with more DM, while a negative coupling (DM $\rightarrow$ DE transfer) produces an opposite effect, similar to a situation with less DM present to drive acoustic oscillations. A $\xi$ parameter different from zero has also the effect of changing the effective EoS parameter $\wla$ of DE, with an ensuing effect on 
the low-$\ell$ part of the spectrum, due to a different contribution to the integrated Sachs-Wolfe (ISW) effect
\cite{He:2010im, Salvatelli:2013wra}. 

A DM/DE coupled scenario also affects physical processes occurring after CMB decoupling, since it alters the background evolution along the whole history of the Universe, it modifies the absolute and relative amount of DM and DE at any epoch, it introduces a non-standard dependence  of the DM density and of its perturbations on time: e.g., 
if DM decays into DE, we might need to start with more DM in the early Universe, corresponding to a stronger clustering
and to an anticipated non-linear regime for the perturbations evolution. It is therefore useful to profit of additional cosmological probes to lift the degeneracy between the DM abundance and the coupling parameter: we will then add to our analysis data from gravitational lensing, clustering and baryonic acoustic oscillations.

\begin{table}[t]
\begin{center}
\begin{tabular}{|c|c|c|c|}
  \hline
  	& \multicolumn{3}{c|}{Prior}		\\ \hline
  Parameter	& \lcdm	& MOD1		& MOD2 	\\ \hline
  $\wla$  	&$-1$	& $[-0.999, -0.1]$& $[-2.5, -1.001]$\\
  $\xi$ 	& $0$	& $[-1, 0]$	& $[0, 0.5]$	\\ \hline
  &no interaction& DM decays into DE & DE decays into DM \\ \hline
\end{tabular}
\caption{Priors on the additional cosmological parameters that define the coupled DM-DE model: the coupling parameter $\xi$ and the DE equation of state parameter $\wla$. Both priors are flat in the listed intervals. The priors for the standard \lcdm~parameters are listed in Tab.~\ref{tab:priorslcdm}.
}
\label{tab:priorscde}
\end{center}
\end{table}

\section{Cosmological Data}
\label{sec:data}
In this section we present the cosmological data we considered for our analyses.

\looseness=-1
We base our analyses on the Cosmic Microwave Background (CMB) data from the recent Planck release \cite{Adam:2015rua}. Specifically, we consider as our minimal data combination
the full temperature autocorrelation spectrum in the range $2\leq\ell\leq2500$ (denoted as ``PlanckTT'') plus the
low-$\ell$ Planck polarization spectra in the range $2 \leq\ell\leq 29$
(denoted as ``lowP'') \cite{Aghanim:2015xee}.
Additionally, we consider and add separately 
the high-$\ell$ Planck polarization spectra in the range
$30\leq\ell< 2500$ (hereafter ``highP'') \cite{Aghanim:2015xee}.

Since the coupling between DE and DM introduces a time dependence in the background evolution of DE and DM 
(see Eqs.~\eqref{eq:coupl_bgDM} and \eqref{eq:coupl_bgDE}), 
it is important to test our theoretical models using data at many different redshift with respect to the CMB measurements.
In particular, in MOD1 we expect an higher amount of DM in the early Universe than in the \lcdm{} model, 
with stronger gravitational effects at earlier evolutionary phases.
On the opposite side, in the MOD2 the amount of DM is smaller in the early Universe 
and the gravitational clustering is reduced until enough DE is decayed into DM.
For these reasons, it is important to consider the effects that observations at various redshift
have on the constraints on the CDE models, in order to attempt distinguishing the different evolution histories.

One of the most important probes of expansion, that are also probes for the existence of DE, 
are the Supernovae (SNe) of type Ia. We consider the luminosity distances of SN Ia 
from the SNLS and SDSS catalogs as re-analyzed in the joint analysis in Ref.~\cite{Betoule:2014frx} (``JLA'' hereafter).

Another probe of the Universe evolution comes from the Redshift Space Distortions (RSD), 
namely distortions of the shape of galaxy clusters in redshift space,
due to peculiar motions of single objects along the line of sight.
The transverse versus line-of-sight anisotropies in redshift space are induced by peculiar motions of the single galaxies
inside a cluster and can provide a way for constraining the growth rate of structures.
This method is under development and 
the detection of RSD is more difficult than a measure of the Baryon Acoustic Oscillations (BAO) features.
See e.g. Ref.~\cite{Percival:2011abc} for a detailed review on RSD.
We include the BAO as determined by
6dFGS \cite{Beutler:2011hx}, 
SDSS-MGS \cite{Ross:2014qpa} and
BOSS DR11 \cite{Anderson:2013zyy},
together with the RSD determinations from \cite{Samushia:2013yga}.
We will refer to the combination of these measurements as to the ``BAO/RSD'' dataset.

The amount of DM affects also the strength of gravitational lensing.
We include information on the power spectrum of the lensing potential reconstructed by Planck 
from the trispectrum measurement \cite{Ade:2015zua} (hereafter ``lens'').
We do not consider weak lensing determinations obtained from the cosmic shear measurements of the CFHTLenS survey
\cite{Heymans:2012gg},
since this put constraints on scales where the non-linear effects are strong. 
Since we want to analyze an interaction model that goes beyond the standard \lcdm~cosmology,
the theoretical description of the non-linear effects can bias our constraints 
when comparing the theoretical predictions to the weak lensing data.
Moreover, at the small scales relevant for the CFHTLenS experiment 
the effects of baryonic feedback and intrinsic alignment can also be important, 
but our knowledge and theoretical description of these effects is quite limited nowadays.
More detailed discussions can be found in Refs.~\cite{Kilbinger:2012qz,Heymans:2013fya,Ade:2015xua}.
We stress that a set of ultra-conservative cuts on the small scales observed by CFHTLenS
is proposed in Refs.~\cite{Ade:2015xua,Ade:2015rim} to exclude  the scales
at which the non-linear evolution have a significant impact.
Even if one applies the ultra-conservative cuts, in the context of the \lcdm~model
the Planck results are in substantial tension with the CFHTLenS results \cite{Raveri:2015maa}.
The tension can be explained invoking the presence
of some unaccounted systematics in the analysis of the experimental data
or an incomplete modeling of the theoretical predictions,
but can also be the result of the existence of new physics beyond the standard model.
However, the importance of the local measurements is nevertheless high, since they provide model-independent results that can be used to constrain the different cosmological models \cite{Verde:2013rgh}. 

A recent analysis \cite{Joudaki:2016mvz} of the CFHTLenS data
that takes into account several astrophysical systematics, however,
shows that the tension between Planck and the cosmic shear measurements
disappears when the systematics are considered jointly.
They find that the two data concordance tests are in agreement,
and that the level of concordance between the two datasets
depends on the exact details of the systematic uncertainties
included in the analysis.
The results of the concordance tests
based on the Bayesian evidence and on information theory
range from decisive discordance to substantial concordance
as the treatment of the systematic uncertainties becomes more conservative.
The least conservative scenario is the one most favored by the cosmic shear data,
but it is also the one that shows the greatest degree of discordance with Planck.
A future, robust result from local measurements that will 
take into account all the possible systematics
will either confirm
the tension with CMB estimates of the cosmological quantities,
probing that the \lcdm~model is incomplete and
possibly suggesting us where to look for new physics,
or confirm that the tension that we observe now is just due to an incomplete
knowledge of some astrophysical phenomenon.

In this respect it is interesting to discuss the results from CFHTLenS
and other experiments that probe the mass distribution at late times.
Among them, we list cluster counts through the Sunyaev-Zel'dovich effect from Planck
\cite{Ade:2013lmv,Ade:2015fva}
(SZ hereafter)
and SPT \cite{Reichardt:2012yj}, from X-ray samples as measured by REFLEX II \cite{Bohringer:2014ooa} 
and
cluster mass distributions at low and high redshifts 
from the Chandra Cluster Cosmology Project \cite{Vikhlinin:2008ym,Burenin:2012uy}.
The measurements of the average matter fluctuations at small scales from these experiments is under discussion,
since the local measurements provide values that can be in tension with Planck determinations from CMB, in the \lcdm~model.
The results are parameterized through the combination $\sigma_8\left(\Omega_M/0.27\right)^\gamma$,
where $\sigma_8$ represents the root-mean-square fluctuations in total matter in a sphere of $8h^{-1}$~Mpc radius,
$\Omega_M$ is the total matter density and $\gamma$ is a parameter that depends on the probed redshift:
the local determinations point towards values for $\sigma_8\left(\Omega_M/0.27\right)^\gamma$ that are lower than the CMB determination.
Differently from the CMB results, however, the local determinations are more likely to suffer some unaccounted systematics:
among the major uncertainties of the local probes there is the overall mass calibration, 
usually quantified through a bias parameter.
The value of the bias parameter is still uncertain, as different indications are found 
when analyzing different samples, as discussed in detail in \cite{Ade:2015xua}.
Again, a clear detection of a preference for a low $\sigma_8$ from the local determinations with respect to the CMB results
would indicate the possible existence of new physics beyond the \lcdm~model.

We do not include in our analyses constraints on the Hubble parameter $H_0$, the expansion rate of the Universe today,
due to the tensions that exist between local determinations and CMB estimates for this observable.
Planck constraints in the context of the \lcdm~model are typically lower than the local measurements. Instead we confront and discuss our results in comparison with the local determination of $H_0$. We will show that MOD2 can
resolve the tension between local and cosmological determinations of both $H_0$ and $\sigma_8$.

To summarize, and to set the reference point for our discussion, we recall that the latest
Planck result in the \lcdm~model is $H_0=67.3\pm1.0\Hou$ 
when using CMB temperature autocorrelation and polarization on large scales only \cite{Ade:2015xua}.
Results on $H_0$ obtained from the CMB are model dependent (they depend on the assumptions on the underlying cosmological model and the type and number of its free parameters), but they do not suffer large systematics.
This is compared with the results obtained by local determinations, 
that in turn can suffer unaccounted systematics but do not depend on the specific cosmological models.
Using the SN Ia detected by HST, with Cepheid-calibrated distances, the authors of Ref.~\cite{Riess:2011yx} found $H_0=73.8\pm2.4\Hou$, 
while using the same SN Ia set with different calibrations for the distance 
it is possible to derive some slightly different value:
for example, a reanalysis of the HST SNe leads to $H_0=70.6\pm3.3\Hou$ when using NGC 4258 as a distance anchor or to
$H_0=72.5\pm2.5\Hou$ when averaging over three different distance-calibration methods 
\cite{Efstathiou:2013via}.
The significance of the tension depends hence on the calibrations. Only $H_0=70.6\pm3.3\Hou$ obtained in Ref.~\cite{Efstathiou:2013via}
is consistent with the CMB result within 1$\sigma$, while the other determinations present some tension.

In our analyses we will explore different combinations of the listed datasets: 
our starting point will be the CMB-only dataset ``PlanckTT+lowP'', 
then we will add one of the other datasets at a time (``highP'', ``lens'',
``JLA'', ``BAO/RSD'') and 
finally we will consider a combination involving all the dataset,
``PlanckTT+lowP + highP + lens + JLA + BAO/RSD'', 
that we will indicate with ``ALL'' for sake of brevity.
For each of these data combinations we will test the three cosmological models (\lcdm, MOD1, MOD2) to investigate
the impact of the coupled DM/DE scenarios.

\section{Results}
\label{sec:results}

\begin{table}[t]                                                                           
\begin{center}                                                                          
\begin{tabular}{|c|c|c|c|}                                                              
\hline
Parameter &	\lcdm	& MOD1	& MOD2\\
\hline
$100\Omega_bh^2$& $2.222\,^{+0.047}_{-0.043}$    & $2.216\,^{+0.046}_{-0.045}$    & $2.226\,^{+0.047}_{-0.046}$    \\
$\Omega_ch^2$	& $0.120\,^{+0.004}_{-0.004}$    & $0.069\,^{+0.051}_{-0.062}$    & $0.133\,^{+0.018}_{-0.015}$    \\
$100\theta$	& $1.0409\,^{+0.0009}_{-0.0009}$ & $1.0441\,^{+0.0051}_{-0.0037}$ & $1.0402\,^{+0.0013}_{-0.0013}$ \\
$\tau$		& $0.078\,^{+0.039}_{-0.037}$    & $0.077\,^{+0.039}_{-0.038}$    & $0.077\,^{+0.039}_{-0.038}$    \\
$n_s$		& $0.965\,^{+0.012}_{-0.012}$    & $0.964\,^{+0.013}_{-0.012}$    & $0.966\,^{+0.013}_{-0.012}$    \\
$\logA$		& $3.089\,^{+0.074}_{-0.072}$    & $3.088\,^{+0.073}_{-0.073}$    & $3.087\,^{+0.073}_{-0.074}$    \\ 	\hline
$\xi$		& $0$                            & $(-0.790,0]$                   & $[0,0.269)$                    \\
$\wla$		& $-1$                           & $[-1,-0.704)$                  & $-1.543\,^{+0.515}_{-0.436}$   \\ 	\hline
$H_0$~[\Hou]	& $67.28\,^{+1.92}_{-1.89}$      & $67.91\,^{+7.26}_{-7.54}$      & $>68.31$                       \\
$\sigma_8$	& $0.830\,^{+0.029}_{-0.028}$    & $1.464\,^{+1.917}_{-0.834}$    & $0.898\,^{+0.163}_{-0.160}$   
\\ 	\hline
\end{tabular}
\caption{Marginalized limits at the 2$\sigma$ C.L.\ for the relevant parameters of this analyses. The results are
obtained with the ``PlanckTT+lowP'' dataset, for the three different models (\lcdm, MOD1 and MOD2).
When an interval denoted with parenthesis is given, it refers to the  2$\sigma$ C.L.\ range starting from the
prior extreme, listed in Tabs.~\ref{tab:priorslcdm} and \ref{tab:priorscde} (and here denoted by the square parenthesis).
}
\label{tab:cmb}
\end{center}
\end{table}

\begin{table}[t]                                                                           
\begin{center}                                                                          
\begin{tabular}{|c|c|c|c|}                                                              
\hline
Parameter &	\lcdm	& MOD1	& MOD2\\
\hline
$100\Omega_bh^2$& $2.229\,^{+0.028}_{-0.028}$    & $2.228\,^{+0.030}_{-0.030}$    & $2.227\,^{+0.031}_{-0.030}$    \\
$\Omega_ch^2$	& $0.119\,^{+0.002}_{-0.002}$    & $0.091\,^{+0.028}_{-0.031}$    & $0.135\,^{+0.014}_{-0.014}$    \\
$100\theta$	& $1.0409\,^{+0.0006}_{-0.0006}$ & $1.0426\,^{+0.0021}_{-0.0018}$ & $1.0400\,^{+0.0010}_{-0.0010}$ \\
$\tau$		& $0.062\,^{+0.025}_{-0.025}$    & $0.063\,^{+0.027}_{-0.026}$    & $0.059\,^{+0.028}_{-0.027}$    \\
$n_s$		& $0.966\,^{+0.008}_{-0.008}$    & $0.966\,^{+0.009}_{-0.009}$    & $0.966\,^{+0.009}_{-0.009}$    \\
$\logA$		& $3.055\,^{+0.045}_{-0.046}$    & $3.058\,^{+0.049}_{-0.049}$    & $3.050\,^{+0.050}_{-0.051}$    \\ 	\hline
$\xi$		& $0$                            & $(-0.463,0]$                   & $0.159\,^{+0.146}_{-0.154}$    \\                                                   
$\wla$		& $-1$                           & $[-1,-0.829)$                  & $(-1.129,-1]$                  \\ 	\hline
$H_0$~[\Hou]	& $67.72\,^{+1.01}_{-0.97}$      & $67.57\,^{+1.81}_{-1.79}$      & $67.83\,^{+1.90}_{-1.75}$      \\
$\sigma_8$	& $0.812\,^{+0.017}_{-0.017}$    & $0.994\,^{+0.283}_{-0.202}$    & $0.749\,^{+0.067}_{-0.061}$    
\\ 	\hline
\end{tabular}
\caption{The same as in Tab.~\ref{tab:cmb} for the analysis on
the ``ALL'' dataset.
}
\label{tab:all}
\end{center}
\end{table}

The results of the analysis for the ``CMB only" and ``ALL" datasets are reported in Tables \ref{tab:cmb} and \ref{tab:all}, respectively. The Tables show the 2$\sigma$ constraints for the relevant parameters of the different analyses. We find that most of the standard cosmological parameters are not sensitive to the coupling in the dark sector and the ensuing results are quite unchanged when moving from \lcdm\  to MOD1 or MOD2: 
the baryon density today $\Omega_b h^2$, the optical depth at reionization $\tau$,
the tilt $n_s$ and amplitude $\logA$ of the power spectrum of scalar perturbations are basically stable
under variation of the cosmological models. Their determination is therefore robust
against modified expansion histories induced by the DM/DE coupling introduced in our modeling.

Slightly larger variations occur for the ratio of the sound horizon to the angular diameter distance at decoupling, $\theta$,
but also in this case the differences between the various models are well inside their mutual 2$\sigma$ limits.
Interestingly, the addition of the external data in the ``ALL'' dataset reduces the uncertainties on various parameters,
but requires a shift towards lower values 
for the optical depth at reionization $\tau$ and the amplitude of the scalar perturbations power spectrum $\logA$.
These parameters suffer of a mild tension in the recent Planck results, as discussed in \cite{Ade:2015xua},
since the analyses that consider the low-$\ell$ temperature spectrum point towards higher values of $\tau$ if compared
to the results obtained from the polarization spectra only. 
If one considers the lensing information and the BAO measurement together with the temperature spectrum,
the results are in good agreement with the indications in favor of a small $\tau$ coming from the Planck polarization spectra.
As the CMB observations constrain the combination $A_s e^{-2\tau}$, a smaller $\tau$ reflects in a smaller $A_s$.

\begin{figure}[t]
  \centering
  \includegraphics[page=1,width=\lenerrbars]{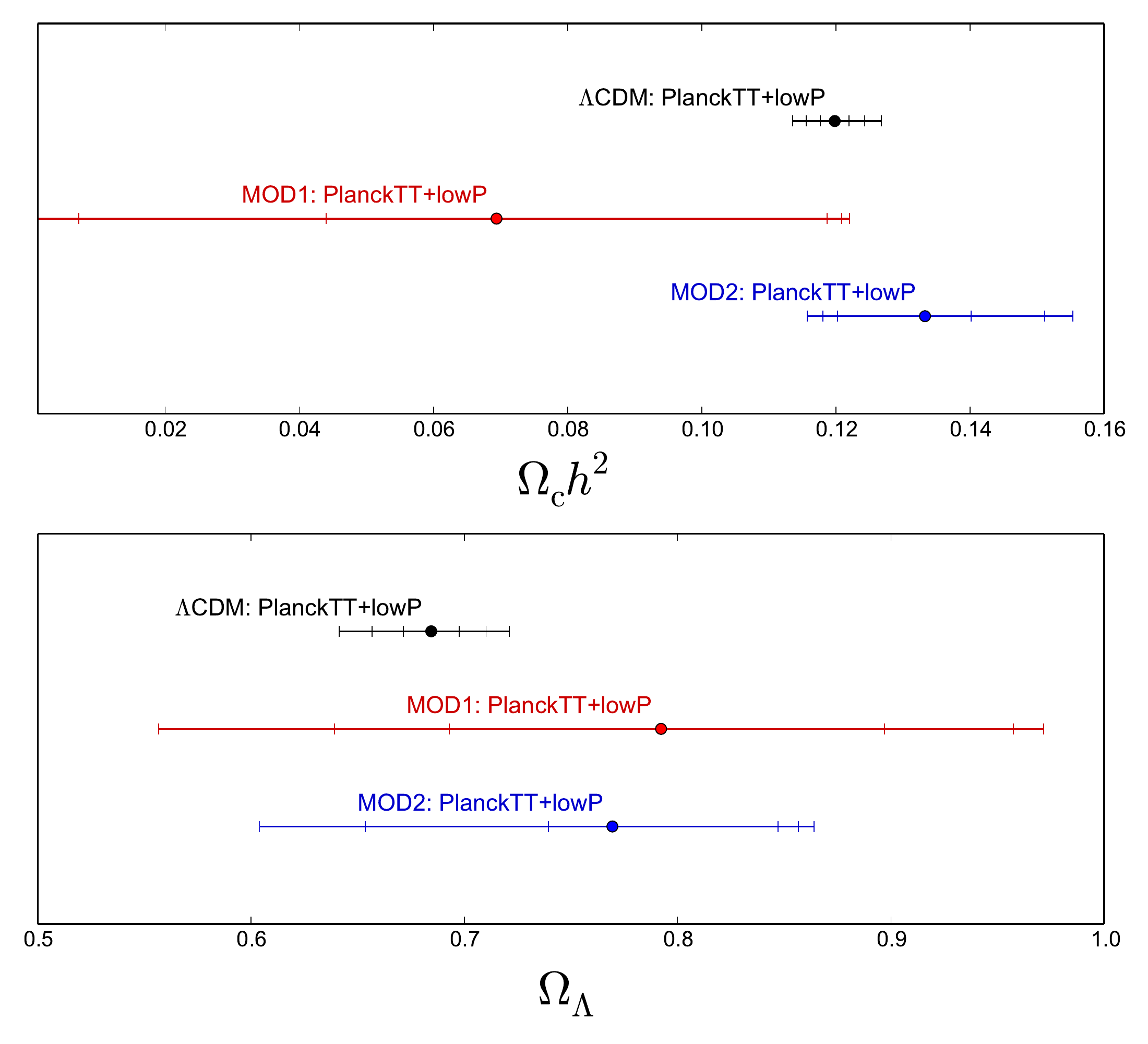}
  \caption{
  Marginalized 1$\sigma$ , 2$\sigma$  and 3$\sigma$ C.L.\ limits on $\Omega_ch^2$ and $\Omega_\Lambda$
   for the ``PlanckTT+lowP'' dataset and for
  the three different cosmological models: \lcdm, MOD1 and MOD2.
  }
  \label{fig:omega1D}
\end{figure}

As expected, there is instead a strong correlation between the coupling parameter $\xi$ and the current DM density $\Omega_ch^2 \propto \rho_c$.
For $\xi<0$ (MOD1), the bigger is the interaction, the smaller is the DM abundance today,
i.e.~more DM decayed into DE during the evolution.
Conversely, for $\xi>0$ (MOD2) a larger DM density is predicted. This is manifest in Tabs.~\ref{tab:cmb} and \ref{tab:all} and in the upper panel in Fig.~\ref{fig:omega1D}, 
where the 1$\sigma$, 2$\sigma$ and 3$\sigma$ C.L.\ intervals for $\Omega_ch^2$ in the different models are shown.
Given a flat Universe (which we assume in our analyses), this turns out in different values for the DE energy density parameter today $\Omega_\Lambda$ in the different models
(see the 1$\sigma$, 2$\sigma$ and 3$\sigma$ C.L.\ bounds for $\Omega_\Lambda$ in the lower panel in Fig.~\ref{fig:omega1D}).

\begin{figure}[t]
  \centering
  \includegraphics[page=2,width=\lenerrbars]{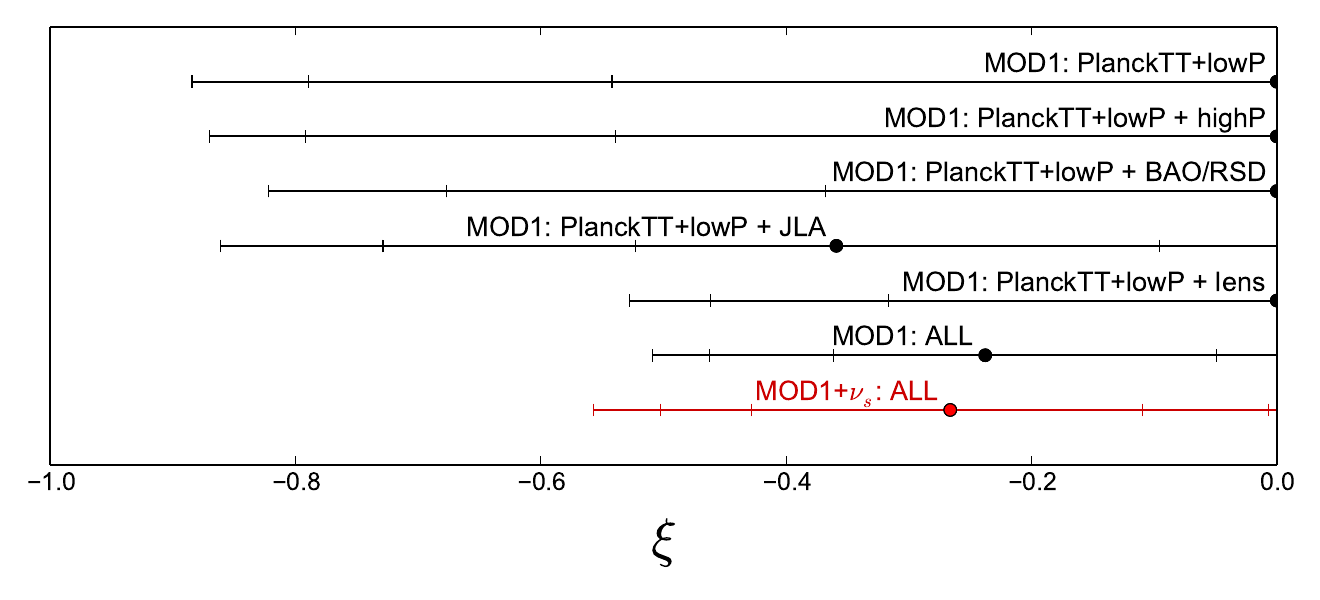}
  \includegraphics[page=3,width=\lenerrbars]{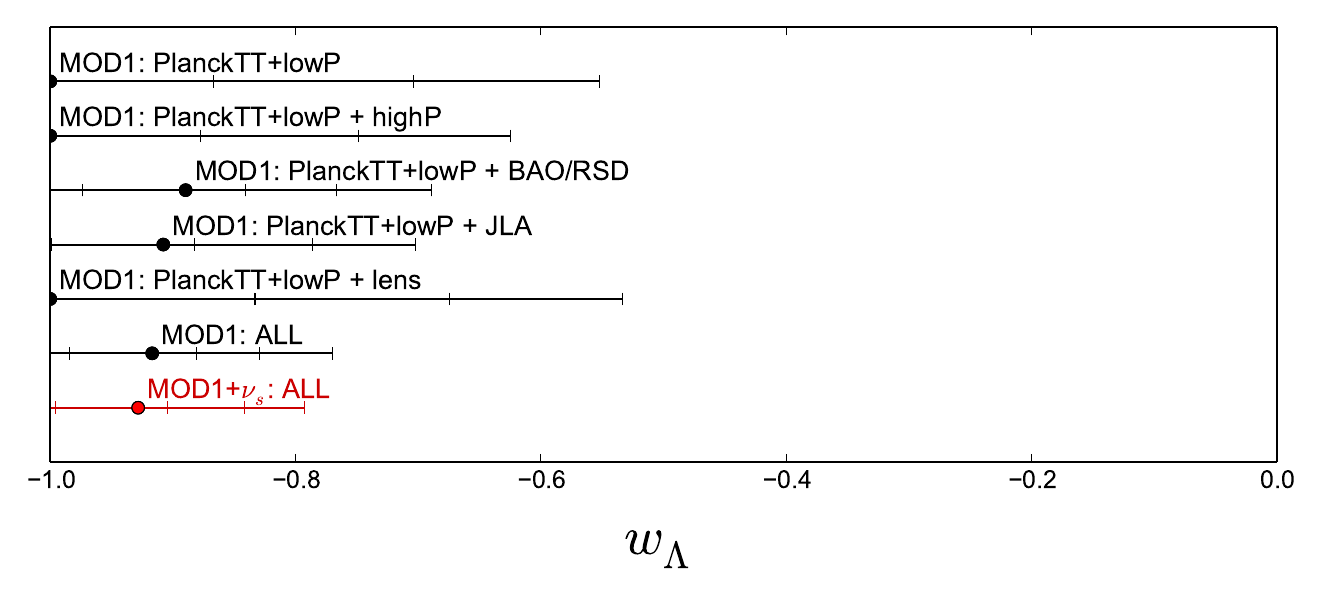}
  \caption{
  Marginalized 1$\sigma$ , 2$\sigma$  and 3$\sigma$ C.L.\ limits for the parameters $\xi$ and $\wla$ in MOD1, for different datasets.
  When the error bar is not visible, it coincides with the limit in the prior, as listed in Tab.~\ref{tab:priorscde}.
  The red point and lines refer to the MOD1\nus~model, discussed in Section~\ref{sec:sterilenuDM}.
  }
  \label{fig:mod1_csi_w}
\end{figure}

\begin{figure}[t]
  \centering
  \includegraphics[page=4,width=\lenerrbars]{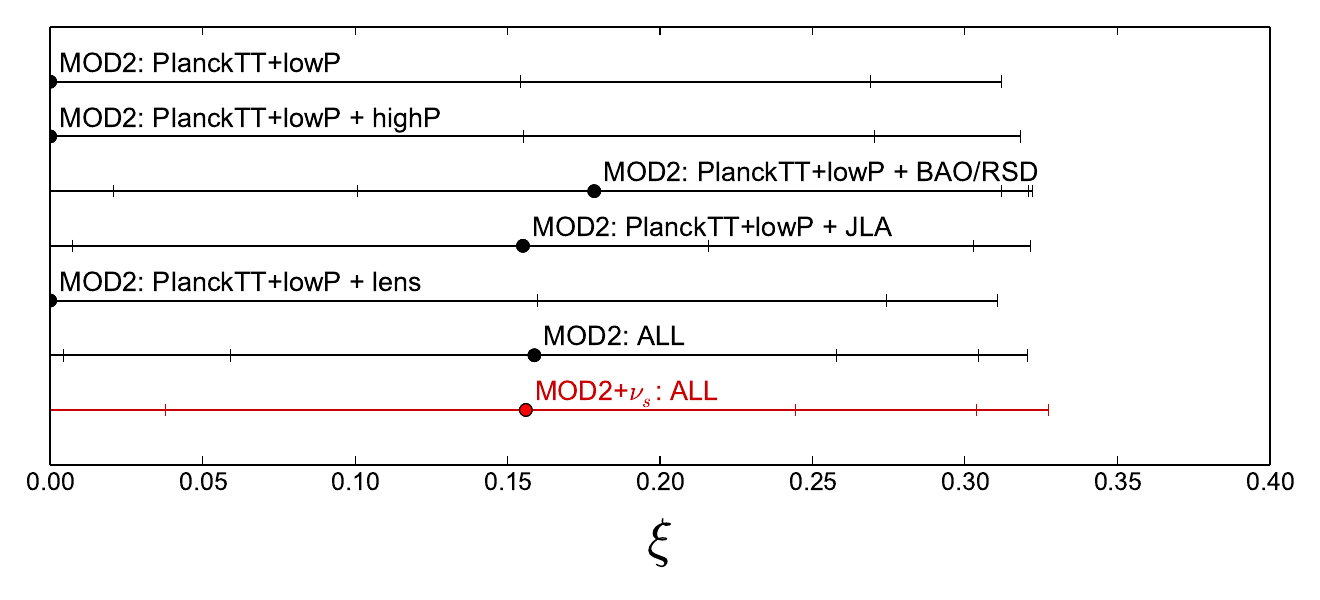}
  \includegraphics[page=5,width=\lenerrbars]{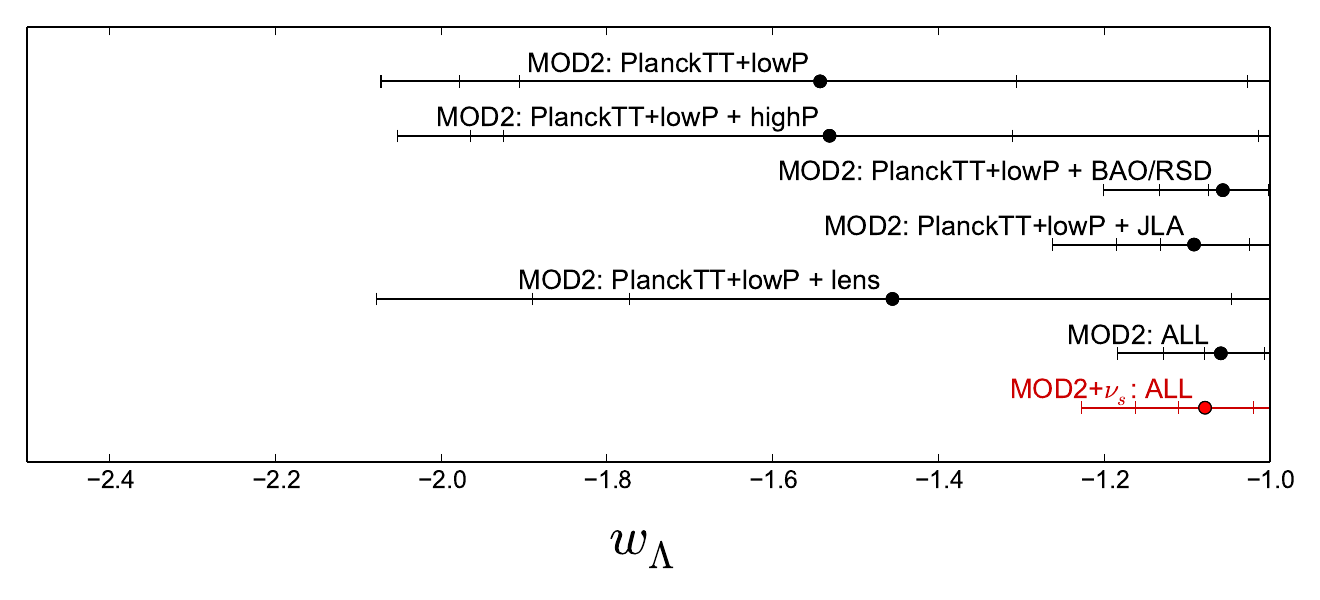}
  \caption{
  Marginalized 1$\sigma$, 2$\sigma$ and 3$\sigma$ C.L.\ limits for the parameters $\xi$ and $\wla$ in MOD2, for different datasets.
  When the error bar is not visible, it coincides with the limit in the prior, as listed in Tab.~\ref{tab:priorscde}.
  The red point and line refer to the MOD2\nus~model, discussed in Section~\ref{sec:sterilenuDM}.
  }
  \label{fig:mod2_csi_w}
\end{figure}

\begin{figure}[t]
  \centering
  \includegraphics[page=2,width=\lencontours]{images/csi_w.pdf}
  \includegraphics[page=1,width=\lencontours]{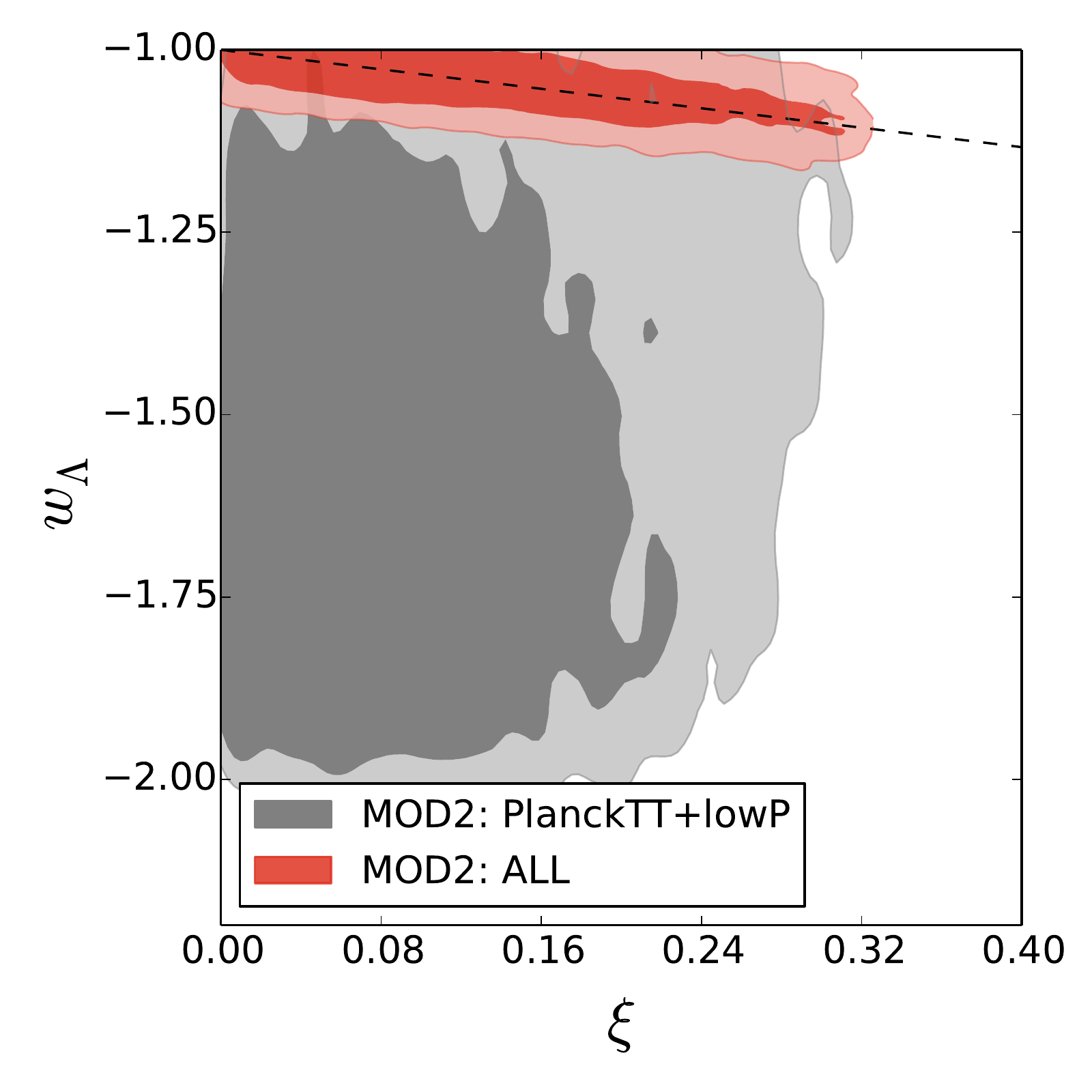}
  \caption{
  Marginalized 1$\sigma$ and 2$\sigma$ C.L.\ allowed regions in the ($\xi$, $\wla$) plane in MOD1 (left) and MOD2 (right), 
  for different datasets. The 
  area below the dashed lines ($\wlae=\wla+\xi/3=-1$) 
  corresponds to an increasing energy density for DE in the future.
  }
  \label{fig:csi_w}
\end{figure}

Figs.~\ref{fig:mod1_csi_w} and \ref{fig:mod2_csi_w} show  the 1$\sigma$, 2$\sigma$ and 3$\sigma$ C.L.\ limits on $\xi$ (upper panels) and $\wla$ (lower panels) obtained with different datasets, 
for both the CDE models MOD1 (Fig.~\ref{fig:mod1_csi_w}) and MOD2 (Fig.~\ref{fig:mod2_csi_w}). 
The constraints are almost insensitive to the addition of the data on CMB polarization at high multipoles (``highP'').
The lensing information, instead, leads to stronger constraints for $\xi$ in MOD1:
as expected, this comes from the bounds on the DM abundance during the expansion history
that are provided by the lensing detection.
Both in MOD1 and MOD2, the addition of the JLA and BAO/RSD dataset lead to stronger constraints on the DE EoS $\wla$, 
that is pushed towards $-1$.
Notice that the analyses constrain
actually 
the effective EoS parameter $\wlae=\wla+\xi/3$: this is the parameter that drives the background evolution in Eq.~\eqref{eq:coupl_bgDE}.
This can also be seen in Fig.~\ref{fig:csi_w}: for both MOD1 (left panel) and MOD2 (right panel) the marginalized regions in the 
($\xi$, $\wla$) plane are well constrained around the $\wlae=-1$ (dashed) line,
thus indicating a preference for a DE energy density that effectively behaves (at the background level) as a cosmological constant, even though at the fundamental level it can interact with DM.

Tab.~\ref{tab:cmb} also shows that the CMB only gives poor constraints on 
$H_0$ and $\sigma_8$ for MOD1 and MOD2.
For the Hubble parameter, this is due to the strong correlation between $H_0$ and the DE EoS parameter:
as we can see in Eq.~\eqref{eq:coupl_bgDE}, when $\wla<-1$ the DE density today is larger
for larger values of $|\wla|$.
Since the Universe is DE dominated at late times, the total energy density $\rho_{\rm{tot}}$ increases with $\rhode$
and consequently the Hubble rate $H\propto\sqrt{\rho_{\rm{tot}}}$ is larger.
When $\wla>-1$, instead, the situation is opposite, and values for $H_0$ lower than the CMB predictions can be found.
The CMB alone, moreover, is not a good way to constrain the DE EoS:
with the introduction of additional data, in particular the BAO/RSD and JLA datasets, the constraints on $\wla$ are much stronger, 
especially in MOD2, and consequently the allowed regions for $H_0$ are better identified.

\begin{figure}[t]
  \centering
  \includegraphics[page=1,width=\lencontours]{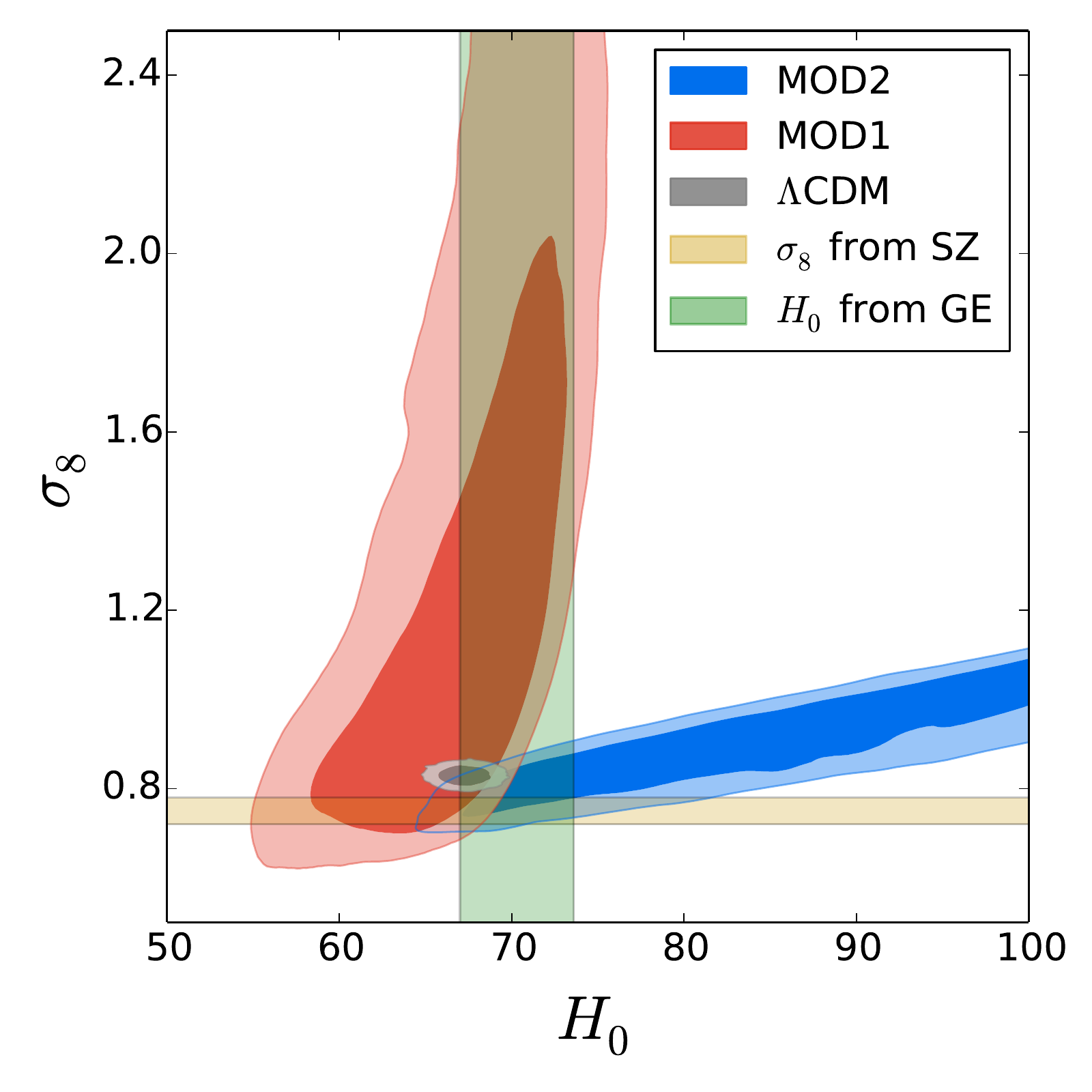}
  \includegraphics[page=2,width=\lencontours]{images/sigma8_h0.pdf}
  \caption{
  Marginalized 1$\sigma$ and 2$\sigma$ C.L.\ allowed regions in the ($\sigma_8$, $H_0$) plane for different models:
  \lcdm~(gray), MOD1 (red) and MOD2 (blue).
  The left panel corresponds to the CMB only dataset ``PlanckTT+lowP'', 
  while the panel on the right refers to the full combination considered here (``ALL'').
  The vertical green band denotes  the interval $H_0=70.6\pm3.3\Hou$ \cite{Efstathiou:2013via} (GE), 
  while the horizontal dark yellow band stands for $\sigma_8=0.75\pm0.03$ \cite{Ade:2013lmv} (SZ).
  }
  \label{fig:sigma8_h0}
\end{figure}

It is interesting to note that MOD1 predicts a value for $\sigma_8$ significantly larger than the \lcdm~prediction
(see both Tab.~\ref{tab:cmb} and Tab.~\ref{tab:all}):
since MOD1 predicts a larger amount of DM in the early Universe, there is more clustering in the primordial Universe, 
that results in an earlier transition to the non-linear evolution and hence to an unavoidably larger value for $\sigma_8$
with respect to the \lcdm~prediction.
Even if the $\sigma_8$ values as determined by local measurements are an underestimate of the true value, nevertheless this fact can be a strong argument against a CDE parameterization through MOD1.
Conversely, in MOD2, the DM abundance is fed by DE as the Universe evolves:
the non-linear evolution starts later and clustering is less prominent, making $\sigma_8$ smaller.
A hint for late-time appearance of DM was found also in a recent study \cite{Pigozzo:2015swa},
thus giving another point in favor of MOD2.
We wish to outline that also  Ref.~\cite{Li:2014cee}, where a CDE scenario with 
$Q\propto\hub\rhode$ is analyzed by adopting  the PPF approach, finds a preference for
a scenario with DE decaying into DM.

In Fig.~\ref{fig:sigma8_h0} we summarize the results on  $H_0$ and $\sigma_8$ in the standard \lcdm\ , MOD1 and MOD2 models.
The left panel refers to CMB data only, while the right panel is for the ``ALL'' datasets.
The two bands show the intervals of the local determinations of $\sigma_8=0.75\pm0.03$ from Planck \cite{Ade:2013lmv},
obtained leaving the mass bias free to vary,
and $H_0=70.6\pm3.3$ \cite{Efstathiou:2013via}.
Both the plots clearly show that MOD1 fails in obtaining high values of $H_0$ accompanied by low $\sigma_8$ values,
and compatible with the local determinations of these parameters.
On the contrary in MOD2
small values of $\sigma_8$ can correspond to large values of $H_0$,
since they are the consequence of the large absolute values allowed for $\wla$ (if the CMB dataset is considered).
The correlation between $\sigma_8$ and $\xi$ is different in the two models:
whereas in MOD1 a larger $\sigma_8$ corresponds to a bigger DM/DE interaction rate, since the larger amount of DM in the early Universe accelerates the evolution of the matter fluctuations at small scales,
MOD2 exhibits an opposite behavior, namely lower values for $\sigma_8$ correspond to a stronger coupling in the dark sector
and possibly to high values of $H_0$, if $\wla$ is large.
In this sense, MOD2 appears to be preferred over MOD1, since in this context the cited tensions of $\sigma_8$ and $H_0$ can be solved.

\section{Sterile neutrinos as stable DM component}
\label{sec:sterilenuDM}

In this Section we extend the dark matter sector to allow
that the total amount of DM is provided by two different species, with only one of them coupled to DE.
In this scenario, the DM is composed by a stable and an interacting fraction, 
with the consequence that only part of the DM can be fed by (or feed) DE during the Universe evolution.
A model with an interacting DM component  combined with a stable one
was studied for instance in Ref.~\cite{Berezhiani:2015yta},
where the authors report a hint for the existence of two separate components.

Among the most investigated DM candidates, sterile neutrinos have been widely studied in the past
(see e.g.~Refs.~\cite{Gariazzo:2015rra,Merle:2013gea,Bergstrom:2014fqa,Merle:2013wta} 
and references therein). We therefore
discuss the possibility that the non-interacting stable DM component is composed by sterile neutrinos
(or any form of matter that behaves in a way similar to sterile neutrinos). 

To include the additional neutrino in the cosmological analysis we use the parameterization presented in
Ref.~\cite{Ade:2013zuv}.
The additional neutrino acts as a relativistic component in the early Universe and gives a contribution
to the effective number of relativistic species \neff:
assuming that the active neutrinos contribute with $\neff^{\rm{sm}}=3.046$ and that there are no other relativistic particles,
the amount of energy density of radiation is given by $\neff>3.046$,
and $\Delta\neff=\neff-\neff^{\rm{sm}}$ measures the effective contribution of the additional neutrino.
This can be defined as \cite{Acero:2008rh}:
\begin{equation}\label{eq:dneff}
  \Delta\neff=
  \left[\frac{7}{8}\frac{\pi^2}{15}{T_{\nu}}^4\right]^{-1}
  \frac{1}{\pi^2}
  \int dp \, p^3 f_{s}(p)\,,
\end{equation}
where $p$ is the neutrino momentum, $f_s(p)$ is its momentum distribution
and $T_\nu$ is the active neutrino temperature.

In the late Universe, when the sterile neutrino becomes non-relativistic, it starts behaving as a dark matter component.
The physical mass $m_s$ is not the most convenient way to describe the sterile neutrino contribution, 
since it enters only into the equation of the energy density
together with the momentum distribution of the neutrino \cite{Acero:2008rh}:
\begin{equation}
  \Omega_{s} h^2
  =
  \frac{h^2}{\rho_c}
  \frac{m_{s}}{\pi^2}
  \int dp \, p^2 f_{s}(p)\,,
\end{equation}
where $\rho_c$ is the critical energy density.
Since the energy density depends on the momentum distribution function and hence on the details of the production of the sterile neutrino in the early plasma,
that in turn depend on the specific underlying neutrino model and properties,
it is more convenient to use instead the effective mass $\meff{s}$, defined as:
\begin{equation}
  \meff{s} = 94.1 \, \mathrm{eV} \, \Omega_s h^2\,.
\end{equation}
If the sterile neutrino is thermally produced with a temperature $T_s$ that is different from that of the active neutrinos,
we have $f(p)= 1/(1+e^{p/T_s})$
and
the two masses are related by $\meff{s} =\Delta\neff^{3/4} m_{s}$.
Differently from Ref.~\cite{Ade:2015xua}, here we do not put constraints on the physical mass of the sterile neutrino,
since we are particularly interested in the degeneracy between $\meff{s}$ and the DM energy density $\Omega_ch^2$.
For both $\neff$ and $\meff{s}$ we adopt flat priors in the intervals listed in Tab.~\ref{tab:priorssterilenu}.

\begin{table}[t]
\begin{center}
\begin{tabular}{|c|c|c|}
  \hline
  	& \multicolumn{2}{c|}{Prior}	\\ \hline
  Parameter	& \lcdm	& $\nu_s$	\\ \hline
  $\meff{s} (eV)$  	& 0	& [0,15]	\\
  $\neff$ 	& 3.046	& [3.046, 6]	\\ \hline
\end{tabular}
\caption{Priors on the neutrino parameters $\meff{s}$ and $\neff$. The priors are assumed flat.}
\label{tab:priorssterilenu}
\end{center}
\end{table}

\begin{figure}[t]
  \centering
  \includegraphics[page=1,width=\lencontours]{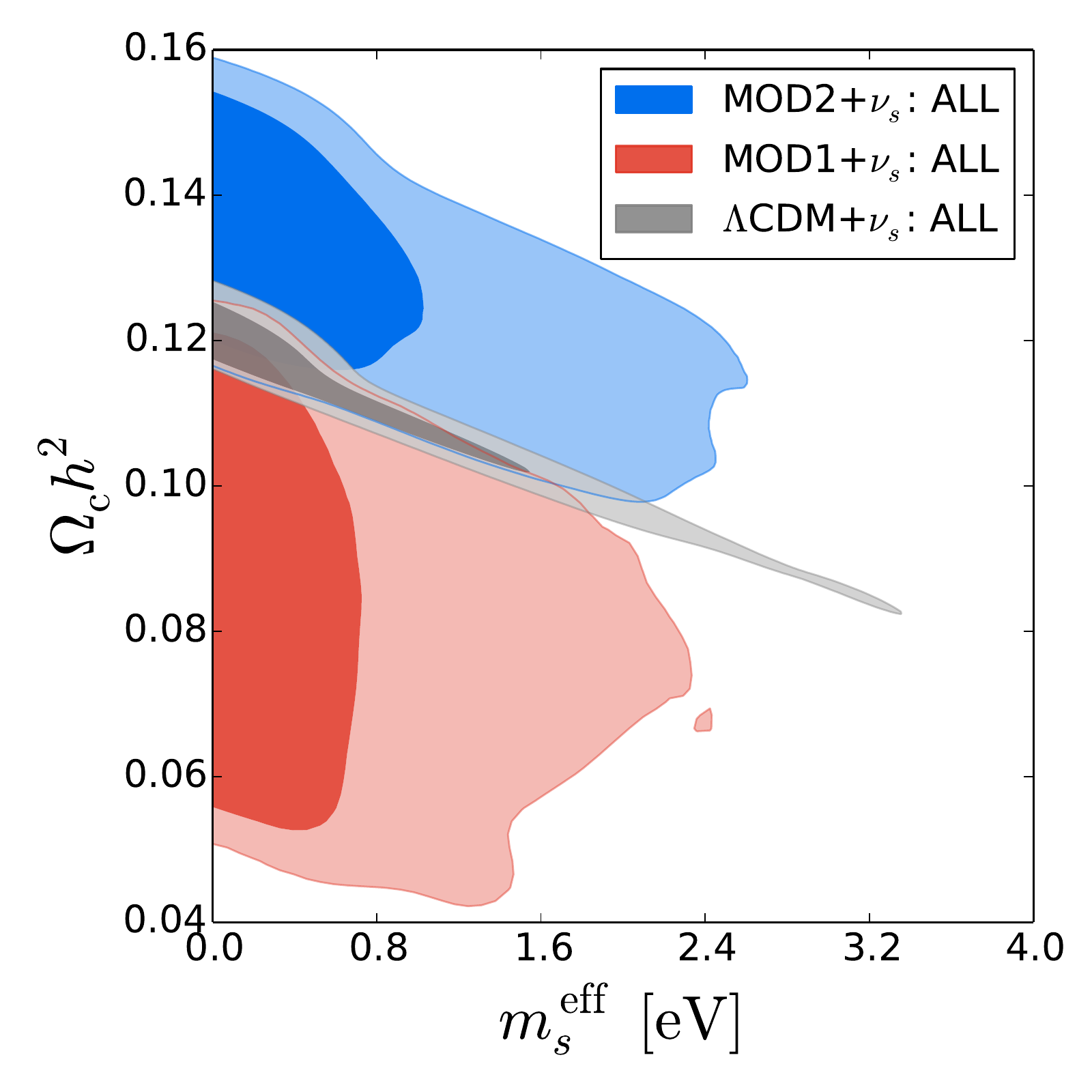}
  \caption{
  Marginalized 1$\sigma$  and 2$\sigma$ C.L.\ allowed regions in the ($\Omega_ch^2$, $\meff{s}$) plane for different models:
  \lcdm\nus~(gray), MOD1\nus~(red) and MOD2\nus~(blue),
  obtained with the full data combination considered here (``ALL'').
  }
  \label{fig:lsn_omc_meff}
\end{figure}

We study the constraints on the sterile neutrino properties using only the full data combination ``ALL'',
that gives the strongest constraints on the CDE models.
We show the results obtained in the \lcdm\nus, MOD1\nus~and MOD2\nus~models in Tab.~\ref{tab:lsn_all} for all the relevant parameters: a comparison with Tab.~\ref{tab:all} shows that the inclusion of an additional neutrino does not change significantly the constraints on the various
parameters, although a small shift down in the baryon density and a small shift up in $\Omega_ch^2$, with an increase of the error bars, can be observed.  These results can be traced to the fact that the sterile neutrino acts as a massive component in the late Universe and it contributes to the total amount of matter with $\Omega_s h^2\propto\meff{s}$ and is therefore degenerate with DM. This degeneracy is clear in Fig.~\ref{fig:lsn_omc_meff}, where a higher DM energy density corresponds to a lower $\meff{s}$, for all the models. The variations on the reconstructed parameters, as compared to the case where the sterile neutrino is not present, are nevertheless well inside 1$\sigma$.

Constraints on the parameters \neff~and \meff{s} are similar in the different models, with only very marginal differences:
this means that the properties of the $\nu_s$ component of DM are not degenerate with the coupling in the dark sector
and the neutrino constraints are robust against the introduction of the new interaction.
In parallel, also the constraints on the coupling parameter $\xi$ and on the DE EoS parameter $\wla$
are largely insensitive to the presence of 
an additional neutrino.
The 1$\sigma$, 2$\sigma$ and 3$\sigma$ C.L.\ limits on these parameters are  plotted in red in Figures~\ref{fig:mod1_csi_w} and \ref{fig:mod2_csi_w} for MOD1 and MOD2 respectively.

\begin{figure}[t]
  \centering
  \includegraphics[page=1,width=\lencontours]{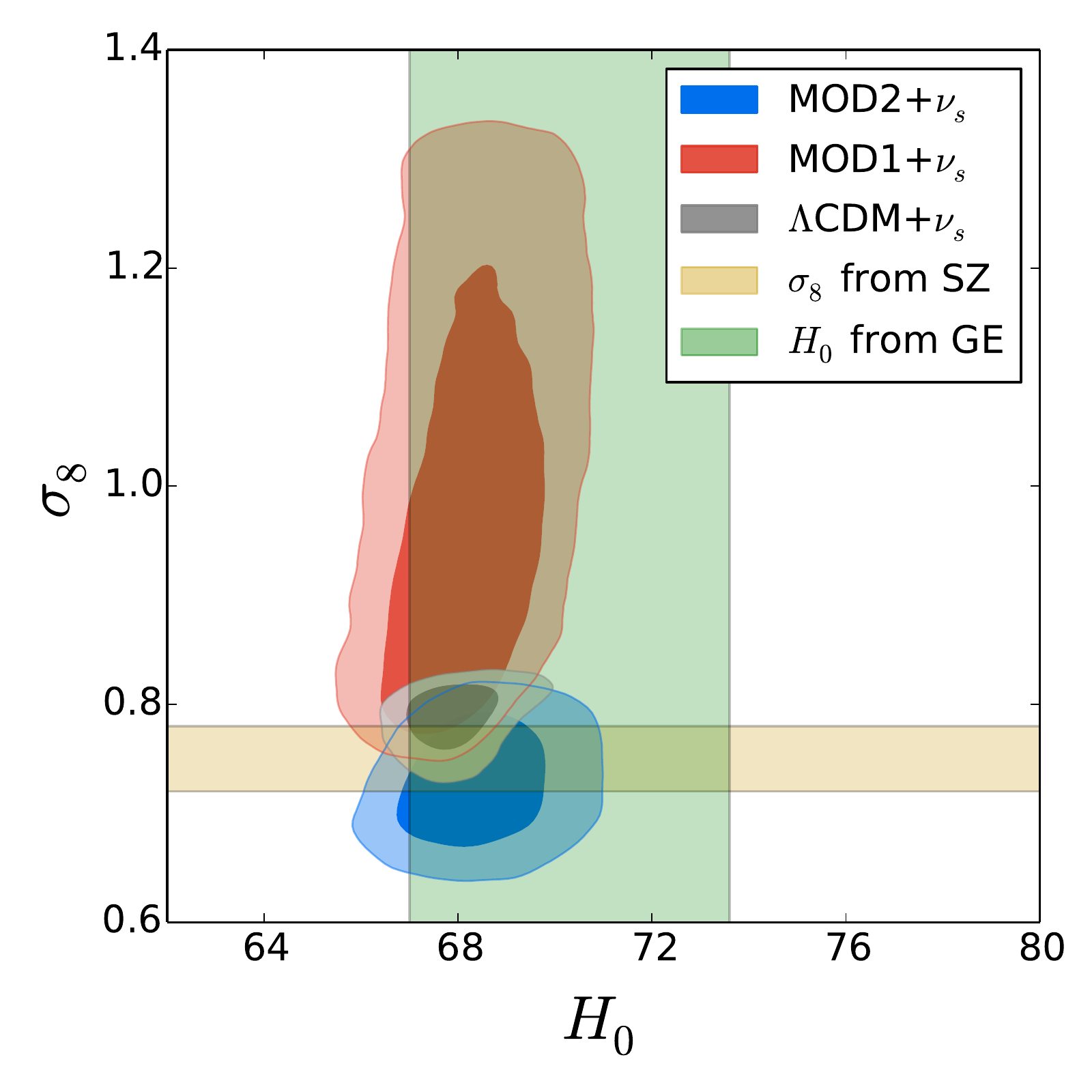}
  \caption{
  Marginalized 1$\sigma$ and 2$\sigma$ C.L.\ allowed regions in the ($\sigma_8$, $H_0$) plane for different models:
  \lcdm\nus~(gray), MOD1\nus~(red) and MOD2\nus~(blue),
  obtained with the full data combination considered here (``ALL'').
  The vertical green band denotes  the interval $H_0=70.6\pm3.3\Hou$ \cite{Efstathiou:2013via} (GE), 
  while the horizontal dark yellow band stands for $\sigma_8=0.75\pm0.03$ \cite{Ade:2013lmv} (SZ).
  }
  \label{fig:lsn_h0_sigma8}
\end{figure}

The presence of an additional component that acts as a relativistic particle in the early Universe
and as a non-relativistic one in the late Universe
gives a suppression in the clustering, due to the free-streaming effect,
and an increase of the Hubble parameter, due to the necessity
of increasing both the DM and DE energy densities in the Universe to avoid
a shift of the matter-radiation equality and of the coincidence time.
As a consequence, the inclusion of the sterile neutrino shifts the predictions for $H_0$ towards slightly higher values
and lowers those for $\sigma_8$.
In Fig.~\ref{fig:lsn_h0_sigma8} we show the equivalent of Fig.~\ref{fig:sigma8_h0} for the models with the additional neutrino.
The regions are slightly wider than in the case with no additional neutrinos, but overall
there are no significant variations with respect to the right panel of Fig.~\ref{fig:sigma8_h0}.
As a consequence of the lowering of $\sigma_8$, however,
models with the sterile neutrino give a slightly improved compatibility
with the low-$\sigma_8$ measurements.

\begin{table}[t]                                                                           
\begin{center}                                                                          
\begin{tabular}{|c|c|c|c|}                                                              
\hline
Parameter &	\lcdm	& MOD1	& MOD2\\
\hline
$100\Omega_bh^2$& $2.237\,^{+0.034}_{-0.031}$    & $2.237\,^{+0.036}_{-0.032}$    & $2.236\,^{+0.035}_{-0.032}$    \\
$\Omega_ch^2$	& $0.113\,^{+0.014}_{-0.019}$    & $0.083\,^{+0.033}_{-0.032}$    & $0.129\,^{+0.023}_{-0.024}$    \\
$100\theta$	& $1.0408\,^{+0.0006}_{-0.0007}$ & $1.0426\,^{+0.0021}_{-0.0019}$ & $1.0400\,^{+0.0011}_{-0.0010}$ \\
$\tau$		& $0.063\,^{+0.032}_{-0.033}$    & $0.064\,^{+0.034}_{-0.035}$    & $0.060\,^{+0.034}_{-0.035}$    \\
$n_s$		& $0.969\,^{+0.012}_{-0.011}$    & $0.968\,^{+0.013}_{-0.011}$    & $0.968\,^{+0.012}_{-0.011}$    \\
$\logA$		& $3.059\,^{+0.066}_{-0.067}$    & $3.061\,^{+0.068}_{-0.070}$    & $3.054\,^{+0.070}_{-0.069}$    \\ 	\hline
$\xi$		& $0$                            & $-0.266\,^{+0.259}_{-0.236}$   & $[0,0.304)$                    \\                      
$\wla$		& $-1$                           & $-0.928\,^{+0.087}_{-0.072}$   & $(-1.162,-1]$                  \\ 	\hline
$\meff{s}$~[eV]	& $<2.1$                         & $<1.8$                         & $<2.2$                         \\                      
$\neff$        	& $<3.34$                        & $<3.38$                        & $<3.35$                        \\ 	\hline
$H_0$~[\Hou]	& $67.91\,^{+1.33}_{-1.26}$      & $68.23\,^{+2.21}_{-2.00}$      & $68.43\,^{+2.11}_{-2.01}$      \\
$\sigma_8$	& $0.789\,^{+0.039}_{-0.045}$    & $0.988\,^{+0.293}_{-0.214}$    & $0.727\,^{+0.076}_{-0.070}$    \\ 	\hline
\end{tabular}
\caption{Marginalized limits at 2$\sigma$ C.L.\ for the relevant parameters of our analyses, 
obtained with the ``ALL'' dataset for the three different models (\lcdm\nus, MOD1\nus~and MOD2\nus).
When an interval denoted with parenthesis is given, it refers to the  2$\sigma$ C.L.\ range starting from the
prior extreme, listed in Tabs.~\ref{tab:priorslcdm} and \ref{tab:priorscde} (and here denoted by the square parenthesis). 
}
\label{tab:lsn_all}
\end{center}
\end{table}

\section{Conclusions}
\label{sec:disc}
The largest part of the energy density of our Universe is represented by a dark sector,
formed by dark matter and dark energy. The presence of
both these components is known only for their gravitational effects,
but we still ignore if they can be explained in the context of fundamental physics:
while many particle candidates have been proposed for DM, the true nature of DE, and whether it has nothing to do with particle physics, is still unknown. We therefore ignore if these two components possess some kind of interaction between themselves, apart from gravity. While it is natural, at least in first approximation, to assume them as separate non-interacting components, as it is usually done in standard cosmology, nevertheless the existence of a non-gravitational coupling involving DE or DM is in principle an option. 

To investigate this possibility, we have performed cosmological tests of a DM/DE interaction, in a model where DM can  partially transfer its density to DE, or vice-versa. The DM/DE interaction has been introduced phenomenologically through an interaction term $Q=\xi H\rhode$ \cite{Zimdahl:2001ar,Salvatelli:2013wra, Costa:2013sva} in the energy conservation equations,
where the dimensionless parameter $\xi$ encodes the coupling strength and the direction of energy flow: from DM to DE for $\xi>0$, from DE to DM for $\xi<0$.

The datasets used to constrain the model have been:
CMB temperature and polarization, gravitational lensing,
supernovae distance calibrations, baryonic acoustic oscillations and redshift space distortions.  The combination of these measurements allows to constrain the evolution of the Universe at different redshifts
and test the DM/DE interaction at different times.

While  Planck observations for CMB temperature and polarization \cite{Adam:2015rua,Aghanim:2015xee} represent our reference datasets, the strongest constraints come with the inclusion of additional information that probe different redshifts:
supernovae data from the joint analysis of Ref. \cite{Betoule:2014frx} strongly constrain the effective DE equation-of-state parameter
$\wlae=\wla+\xi/3$ to be close to $-1$,
while BAO/RSD \cite{Beutler:2011hx,Ross:2014qpa,Anderson:2013zyy,Samushia:2013yga}
data give a mild preference for a non-zero coupling, both for MOD1 and MOD2.
If we consider the derived values of the Hubble parameter $H_0$ and of $\sigma_8$,
we find that MOD1 (a model that was studied e.g in Refs.~\cite{Costa:2013sva, Salvatelli:2013wra}),
increases the tension with the low-redshift measurements of $H_0$ from HST \cite{Riess:2011yx, Efstathiou:2013via}
and, more significantly, with the Planck SZ cluster counts, CFHTLenS and other local determinations of $\sigma_8$
\cite{Ade:2013lmv, Ade:2015fva, Reichardt:2012yj,Vikhlinin:2008ym,Bohringer:2014ooa, Heymans:2012gg}.
The reason is that in MOD1 a higher amount of DM in the early Universe is required to allow for the survival of a necessary amount of DM today:
this larger DM amount in the early phases increases the clustering effect and 
forces the non-linear evolution to occur earlier. 
On the contrary, in MOD2, $\sigma_8$ is smaller than in the \lcdm~model, as a consequence of the DM to DE transfer, and cosmological determinations of $H_0$ and $\sigma_8$ are better reconciled with low-redshift probes.

We studied also the possible presence of a sterile neutrino as an additional and stable dark matter component \cite{Merle:2013gea,Gariazzo:2015rra}.
In this case we find that the sterile neutrino parameters are completely insensitive to
the parameters of the CDE model and the constraints are basically the same for the \lcdm\nus, the MOD1\nus~and the MOD2\nus~models.

In conclusion, a coupled DM/DE cosmology is a viable option, compatible with a large host of cosmological data. Moreover, a model where
DE feeds DM 
during the evolutionary history of the Universe can help solving the small tensions that currently exist between different high- and low-redshift observations in the context of the \lcdm~model, therefore providing an interesting new opportunity of investigation for models of the dark sectors of the Universe.


\begin{acknowledgments}
This work is supported by the research grant {\sl Theoretical Astroparticle Physics} number 2012CPPYP7 under the program PRIN 2012 funded by the Ministero dell'Istruzione, Universit\`a e della Ricerca (MIUR), by the research grants {\sl TAsP (Theoretical Astroparticle Physics)} and {\sl Fermi} funded by the Istituto Nazionale di Fisica Nucleare (INFN), and by the {\sl Strategic Research Grant: Origin and Detection of Galactic and Extragalactic Cosmic Rays} funded by Torino University and Compagnia di San Paolo.
\end{acknowledgments}

\bibliographystyle{utcaps}
\bibliography{cde}

\end{document}